\documentclass[conference]{IEEEtran}
\IEEEoverridecommandlockouts
\usepackage[style=ieee]{biblatex}
\usepackage{amsmath,amssymb,amsfonts}
\usepackage{algorithm}
\usepackage{algpseudocode}
\usepackage{graphicx}
\usepackage{textcomp}
\usepackage{xcolor}
\usepackage{tablefootnote}

\def\BibTeX{{\rm B\kern-.05em{\sc i\kern-.025em b}\kern-.08em
    T\kern-.1667em\lower.7ex\hbox{E}\kern-.125emX}}
\begin{document}

\title{Pianist Identification Using\\Convolutional Neural Networks
\thanks{This work is supported by the UKRI Centre for Doctoral Training in Artificial Intelligence and Music (EP/S022694/1). J.Tang is a research student supported jointly by the China Scholarship Council and Queen Mary University of London. G.~Wiggins received funding from the Flemish Government under the ``Onderzoeksprogramma Artificiële Intelligentie (AI) Vlaanderen''.}
}

\author{\IEEEauthorblockN{1\textsuperscript{st} Jingjing Tang}
\IEEEauthorblockA{\textit{Center for Digital Music} \\
\textit{Queen Mary University of London}\\
London, UK \\
jingjing.tang@qmul.ac.uk}

\and
\IEEEauthorblockN{2\textsuperscript{nd} Geraint Wiggins}
\IEEEauthorblockA{\textit{Computational Creativity Lab}\\\textit{Vrije Universiteit Brussel, Belgium} \&\\
\textit{Queen Mary University of London, UK} \\
geraint.wiggins@vub.be}
\and
\IEEEauthorblockN{3\textsuperscript{rd} Gy\"orgy Fazekas}
\IEEEauthorblockA{\textit{Center for Digital Music} \\
\textit{Queen Mary University of London}\\
London, UK \\
george.fazekas@qmul.ac.uk}
}
\maketitle
\begin{abstract}
This paper presents a comprehensive study of automatic performer identification in expressive piano performances using convolutional neural networks (CNNs) and expressive features. Our work addresses the challenging multi-class classification task of identifying virtuoso pianists, which has substantial implications for building dynamic musical instruments with intelligence and smart musical systems.  Incorporating recent advancements, we leveraged large-scale expressive piano performance datasets and deep learning techniques. We refined the scores by expanding repetitions and ornaments for more accurate feature extraction. We demonstrated the capability of one-dimensional CNNs for identifying pianists based on expressive features and analyzed the impact of the input sequence lengths and different features. The proposed model outperforms the baseline, achieving 85.3\% accuracy in a 6-way identification task. Our refined dataset proved more apt for training a robust pianist identifier, making a substantial contribution to the field of automatic performer identification. Our codes have been released at \url{https://github.com/BetsyTang/PID-CNN}.
\end{abstract}


\section{Introduction}\label{intro} 
 Performers, with their individual phrasing, dynamics, and interpretive choices, bring their personal artistry to each piece they play, resulting in distinguishable styles. Researchers who focus on studying expressive musical performances have been investigating computational models for performer identification \cite{stamatatos2005automatic, ramirez2010automatic, ce84fcd7918944e389a7e65b01fe6a9a, zhao2021violinist, syed2023hipi}. A reliable pianist identifier holds great potential for not only studying the styles of different performers, but also various applications in music education, music information retrieval and smart musical instruments~\cite{turchet2019smi}. As an illustration, a pianist identification model could aid piano students wishing to emulate the performances of virtuoso pianists. With an upsurge in embedded devices, the vision of smart musical systems—ones that can discern different performers or styles and provide real-time feedback or adjustments—becomes closer to reality. Imagine a smart piano capable of tailoring its settings to mirror the nuances of iconic pianists, or a wearable accessory that offers pianists instant feedback, juxtaposing their performance against the masterpieces of legendary artists. Networked musical instruments could use style information or the features extracted by the proposed system in educational, retrieval or networked performance contexts, similar to those proposed by Turchet et.al. in~\cite{turchet2020cludsmart}. These groundbreaking applications will not only resonate with the principles of Internet of Musical Things (IoMusT)~\cite{8476543} and the Internet of Audio Things (IoAuT)~\cite{9099251} but also elevate their potential, transforming basic devices into dynamic musical instruments with intelligence in the context of the Internet of Sounds (IoS)~\cite{turchet2023internet}.
 
Automatic performer identification is usually regarded as a multi-class classification task where the system is designed to infer the performer of the given music performance. Early studies \cite{stamatatos2005automatic, ramirez2010automatic} mainly applied traditional machine learning algorithms such as K-means clustering, decision trees, and discriminant analysis to this task. More recent research \cite{zhao2021violinist, ce84fcd7918944e389a7e65b01fe6a9a} calculated the KL-divergence between performers' feature distributions and identified performer by performing similarity estimation based on the KL-divergence. Zhao et al.~\cite{9909590} utilised transfer learning for classifying violinists, adopting pre-trained models for music tagging and singer identification. With the emergence of large-scale expressive piano performance datasets \cite{zhang2022atepp, foscarin2020asap}, two projects \cite{9909590, syed2023hipi, zhang2022atepp} recently applied deep learning techniques to pianist identification task. Rafee et al.~\cite{syed2023hipi} proposed a RNN-based hierarchical neural network for pianist identification. Zhang et al.~\cite{zhang2022atepp} has applied convolutional neural networks (CNNs) to a 16-way pianist identification task, achieving less than 50\% accuracy. However, this work paid insufficient attention to extracting expressive features which have been proven effective for deep neural networks that model expressiveness and performance styles of pianists \cite{Jeong2019VirtuosoNetAH, syed2023hipi}.

This paper details our exploration of the potential of CNNs in identifying virtuoso pianists using various expressive features. We obtained a subset consisting of both performance and score midis from the ATEPP dataset, refining the scores by extending the repetitions and ornaments in the corresponding midis, thus generating the most comprehensive and accurate dataset currently available for pianist identification. We conducted experiments to investigate the effectiveness of different expressive features and the impact of input sequence lengths. The proposed one-dimensional CNN surpassed the baseline model \cite{syed2023hipi}, attaining an 85.3\% accuracy for a 6-way identification task. In addition, our dataset was shown to be more suitable for training a robust pianist identifier compared to the one proposed previously \cite{syed2023hipi}.

The rest of this paper is organised as follows: Section \ref{method} elaborates on the methodology, providing details of the dataset, the feature extraction process, and the model architecture. Section \ref{experiments} outlines the experiment set-ups employed for model training. Section \ref{results} discusses the experiment results and the ensuing discussions. Lastly, Section \ref{conclution} concludes the paper.
\section{Methodology}\label{method}
\subsection{Dataset}
As discussed by Rafee et al.~\cite{syed2023hipi}, the lack of large datasets containing multiple performances of the same compositions by different pianists results in the lack of investigation in deep neural networks for pianist identification. However, the recent proposed expressive piano performance midi dataset, ATEPP \cite{zhang2022atepp}, enabled us to create subsets which are balanced in the number of performances for six virtuoso pianists including Alfred Brendel, Claudio Arrau, Daniel Barenboim, Friedrich Gulda, Sviatoslav Richter, and Wilhelm Kempff. In our research, we consider two subsets as shown in the Table \ref{tab:data4pi}:
\begin{itemize}
    \item[(1)] \textit{ID-400}: we created an updated version of the proposed subset by Rafee et al.~\cite{syed2023hipi} by removing corrupted transcription results as well as repeated performances following the latest version of the ATEPP dataset\footnote{https://github.com/BetsyTang/ATEPP}.
    \item[(2)] \textit{ID-1000}: we chosen a larger subset containing more compositions and performances by the same pianists to increase robustness and verify the capability of our model.
\end{itemize}

\begin{table}[!hbt]
    \centering
    \caption{Number of performances for each pianist and other statistics in selected subsets}
    \begin{tabular}{ccc}
    \hline
    \textbf{Datasets}&\textbf{\textit{ID-400}  \cite{syed2023hipi}}&\textbf{\textit{ID-1000}}\\
    \hline
    Sviatoslav Richter&77&194\\
Alfred Brendel&77&192\\
Friedrich Gulda&77&170\\
Claudio Arrau&77&163\\
Daniel Barenboim&73&150\\
Wilhelm Kempff&76&132\\
\hline
Total Performances&457&1001\\
Movements&36&40\\
Hours&45.9&109.4\\
\hline
    \end{tabular}
    \label{tab:data4pi}
\end{table}

All movements in both subsets are by Beethoven or Mozart. Each movement corresponds to at least one performance by each pianist, making it possible to compare the differences in performance style of each individual performer. In order to maintain similar data distributions in training, validation, and testing sets, we divided the datasets alongside the number of performances of a composition by each pianist. To achieve a 8:1:1 train-valid-test split, we followed the Algorithm \ref{alg:data} to assign performances to \textit{Train}, \textit{Valid} and \textit{Test} subsets. The Algorithm \ref{alg:data} is designed to guarantee that each split contains at least one performance of the composition by a performer, especially when there are fewer than 10 performances by that performer.
\begin{algorithm}
\caption{Data Splitting}\label{alg:data}
\begin{algorithmic}
\State \# Let $C$ be the set of compositions, $P$ be the set of pianists. 
\State \# $\mathit{Info}$ returns composition and pianist of a performance $i$. 
\State \# $\mathit{Count}$ gives the number of performances in a set $S$. 
\State \# $\mathit{RandomSplit}$ randomly splits a set $S$ of size $n$ into subset $a$ of size $rn$ and subset $b$ of size $(1-r)n$, where $r \in [0, 1)$.
\State \# $\mathit{Random}$ generates a number between $[0, 1)$ following the uniform distribution.
\State \# $\gets$ means "assigned to".
\For{($c$, $p$) in ($C$, $P$)}
\State $n = \mathit{Count}(I)$, where if $i \in I$, $\mathit{Info} (i) = (c,p)$
\If {$n \leq 1$} $Train \gets I$
\ElsIf {$n=2$} $a, b = \mathit{RandomSplit}(I, r=1/n)$
\State $m = \mathit{Random}()$, $Train \gets b$
\If {$m \leq 0.5$} $Valid \gets a$
\ElsIf {$m > 0.5$} $Test \gets a$
\EndIf
\ElsIf {$3\leq n\leq9$} $a, b = \mathit{RandomSplit}(I, r=\frac{1}{n})$
\State $b, c = \mathit{RandomSplit}(b, r=\frac{1}{n-1})$
\State $Valid \gets a$, $Test \gets b$, $Train \gets c$
\ElsIf{$10\leq n$} $a, b = \mathit{RandomSplit}(I, r=\frac{4}{5})$
\State $b, c = \mathit{RandomSplit}(b, r=\frac{1}{2})$
\State $Train \gets a$, $Valid \gets b$, $Test \gets c$
\EndIf
\EndFor
\end{algorithmic}
\end{algorithm}

\subsection{Score and Performance Alignment}
Inspired by previous research \cite{stamatatos2005automatic, ce84fcd7918944e389a7e65b01fe6a9a, syed2023hipi} focusing on pianist identification, we used an alignment algorithm proposed by Nakamura et al.~\cite{Nakamura2017PerformanceED} to establish correspondences between performance midi data and score midi data, which allowed us to extract performance-related features. While the algorithm exhibited promising results in most cases, it demonstrated limited capability in handling annotated repetitions and ornaments found in the scores. To address this limitation, we manually expanded the repetitions and added ornament notes to the score midi files, thereby enhancing the accuracy of the alignment results. The improved alignment results more accurately captured the nuances of performances, aiding in distinguishing among performers.

After performing the alignments, we proceeded to filter out two types of discrepancies: \textit{missing notes} (representing notes present in the scores but not successfully aligned to performances) and \textit{extra notes} (representing notes present in performances but not successfully aligned to scores). Then we quantified the extent of information loss caused by the alignment algorithm for each performance, as captured by Equation~\ref{equ: infoloss}:
\begin{equation}
\label{equ: infoloss}
\textit{Loss of Information} = \frac{N_e}{N_p} \times 100\%,
\end{equation}
where $N_e$ denotes the number of extra notes and $N_p$ refers to the total number of notes in the performance. 
\begin{figure}[!hbt]
  \centering
\includegraphics[width=.95\linewidth]{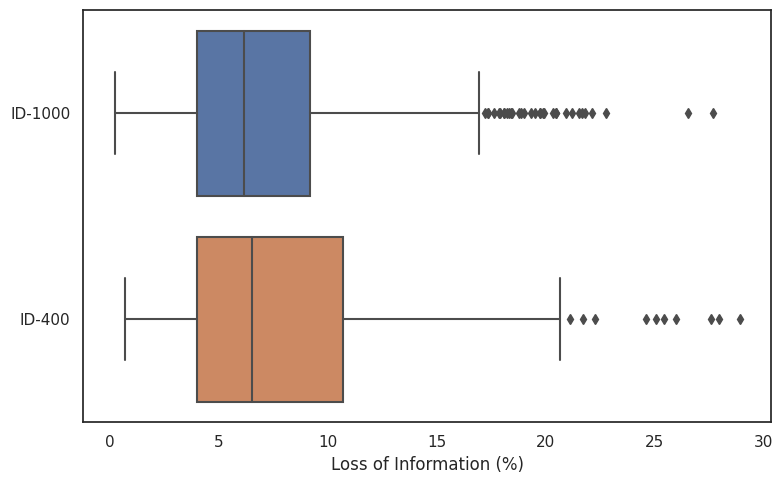}
  \caption{Boxplots of information loss caused by the alignment process in \textit{ID-400} and \textit{ID-1000} datasets}
  \label{fig:infoloss}
\end{figure}
The distributions of information loss in the datasets \textit{ID-400} and \textit{ID-1000} are presented in Fig.~\ref{fig:infoloss}. Our analysis reveals that more than 95\% of performances in both datasets exhibit less than 15\% information loss.

\subsection{Feature Extraction}\label{sec:feature}
After aligning the performances and scores, we extracted input features following the process outlined in the study by Rafee et al.~\cite{syed2023hipi}. We derived deviations between the scores and performance for note-wise features, encompassing aspects such as timing and velocity. Beyond considering feature deviations, we also incorporated the original note-wise features as part of our input data. A full list of features used for our experiments are summarised in the Table \ref{tab:features}. Two note-level features are defined as follows: \textit{Inter-onset Interval} (IOI), representing the temporal duration between the onset times of two consecutive notes, and \textit{Offset Time Duration} (OTD), signifying the time interval between the offset time of a note and the onset time of its subsequent note. To process the features into suitable input for our model, we organized them into sequences, preserving the order of the notes. These sequences were then stacked together to create the final input. The resulting shape of the input would be (\textit{batch size}, \textit{sequence length}, \textit{number of features}), as shown in the Fig.~\ref{fig:model} at the left side.

To examine the performance of our model under circumstances of limited information, we divided the sequences into segments of varying lengths respectively. This allowed us to gauge the model's capacity to manage scenarios with limited data availability, detailed further in Section \ref{results}.

\begin{table}[!hbt]
    \centering
     \label{tab:features}
    \caption{Summary of features used in model training}
    \begin{tabular}{cc}
    \hline
    \textbf{Note-wise Feature}&\textbf{Deviations}\\
    \hline
    Pitch&No\\
    Velocity&Yes\\
Onset time&Yes\\
Offset time&Yes\\
Duration&Yes\\
Inter-onset Interval (IOI)&Yes\\
Offset Time Duration (OTD)&Yes\\
\hline
    \end{tabular}
\end{table}
\subsection{Model Architecture}
\begin{figure}[!hbt]
  \centering
\includegraphics[width=\linewidth]{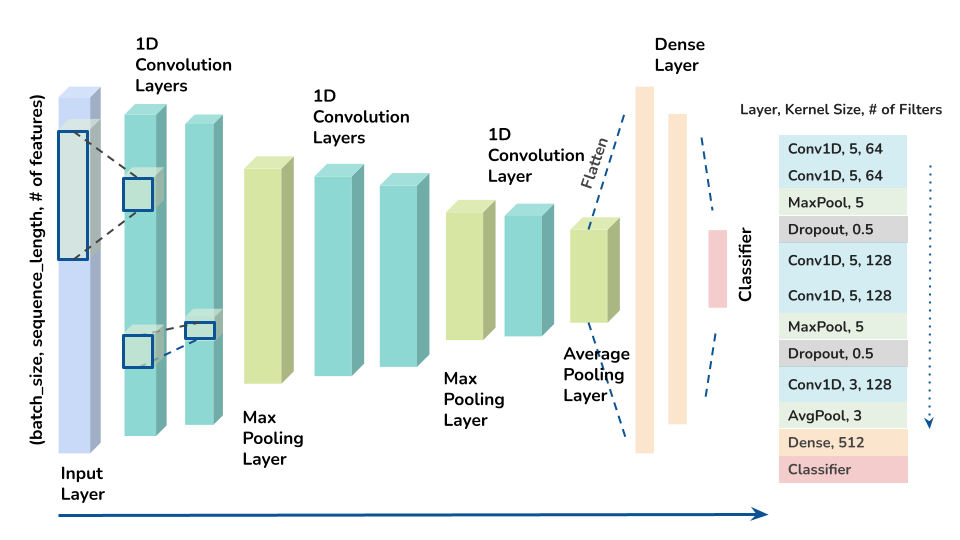}
  \caption{Model architecture of the proposed one-dimensional CNN}
  \label{fig:model}
\end{figure}

In light of the promising performance demonstrated by Convolutional Neural Networks (CNNs) in various classification tasks across different domains, we proposed a novel one-dimensional CNN model to address the pianist identification task. The architecture was determined through an empirical grid search, focusing on structural hyperparameters such as the number of layers and kernel size. The model architecture, depicted in Fig.~\ref{fig:model}, encompasses five convolutional layers followed by one dense layer, strategically designed to efficiently process the input data. All convolution layers are followed by a ReLU activation and a batch normalization layer. Dropout layers are added in order to avoid overfitting problem. 

\section{Experiments}\label{experiments}
We implemented our model using PyTorch \cite{paszke2019pytorch}, and monitored and recorded the experimental progress through the use of Wandb \cite{wandb}. To achieve optimal model performance, we conducted an extensive hyperparameter tuning process using grid search. We specifically focused on parameters such as learning rate, weight decay, batch size, and the number of training epochs. This process was enhanced by leveraging the powerful capabilities of Wandb Sweeps. Consequently, our model underwent training with a batch size of 16 for a total of 1500 epochs, employing the Adam optimizer with an initial learning rate set to 8e-5 and a weight decay rate of 1e-7.

Our proposed model, which has only 6.1 million trainable parameters, showcasing remarkable efficiency. On average, a single experiment on a GeForce RTX 2080 Ti GPU takes approximately 1.2 hours. This duration stands in stark contrast to the significantly lengthier training times encountered in the context of RNN-based hierarchical models, as proposed by Rafee et al.~\cite{syed2023hipi}. 
\section{Results}\label{results}
To thoroughly evaluate our proposed CNN model in addressing the pianist identification task, we conducted three studies. These studies examined the impacts of variable input sequence lengths, the diverse expressive features, and the datasets on the model's performance. To ensure a reliable assessment of the model, each experiment was repeated three to five times under consistent experimental settings. For a more straightforward comparison with the state-of-the-art \cite{syed2023hipi}, both Study I and II were conducted using the \textit{ID-400} dataset.

\subsection{Study I: Effect of Varying Input Music Sequence Lengths}
The reliable identification of a pianist necessitates stable performance regardless of variations in the length of the musical input. We embarked on a series of experiments using all the features delineated in Section~\ref{sec:feature} to train our model. Experiments were conducted on complete musical pieces and segments of varying lengths, utilizing the \textit{ID-400} dataset. Mean values along with standard deviations pertaining to accuracy and F1-score for each experiment are tabulated in Table~\ref{tab:result1}. As inferred from the outcomes, our model demonstrated uniform high performance when dealing with sequences comprising 1000 notes or less. However, incorporating the full scope of performances substantially bolstered the model's performance as opposed to relying solely on performance segments. Furthermore, our model surpassed the benchmark set by the state-of-the-art RNN-based Hierarchical model \cite{syed2023hipi} when we integrated more features into the training at both piece-wise and segment-wise levels. Our model attained a commensurate level of accuracy when trained with the same number of features as their study.
\begin{table}[!hbt]
    \centering
    \caption{Study I Results on \textit{ID-400} with Varied Input Lengths: Comparing Our Model to the State-of-the-Art \cite{syed2023hipi} at Piece-Wise (Full) and Segment-Wise (1000) Levels}
    \begin{tabular}{ccccc}    
    \hline
    \textbf{Model}&\textbf{Length$^{\star}$}&\textbf{Fea.$^{\dagger}$}&\textbf{Acc. (Std.)}&\textbf{F1 (Std.)}\\
    \hline
    Ours&400&13&0.781 (0.020)&0.768 (0.021)\\
    Ours&600&13&0.762 (0.006)&0.729 (0.014)\\
    Ours&800&13&0.768 (0.019)&0.768 (0.017)\\
    Ours&1000&13&\textbf{0.787 (0.015)}&\textbf{0.781 (0.017)}\\
    Ours&Full&13&\textbf{0.853 (0.033)}&0.804 (0.042)\\
    \hline
    Ours&1000&3&0.720 (0.008)&0.719 (0.007)\\
    Ours&Full&3&0.815 (0.022)&0.806 (0.031)\\
    Hierarchical \cite{syed2023hipi}&1000&3&0.760&0.755\\
    Hierarchical \cite{syed2023hipi}&Full&3&0.847&\textbf{0.845}\\
\hline
    \end{tabular}
    \label{tab:result1}
    \vspace{1mm}
    \raggedright
    \footnotesize{$^{\star}$ The length number refers to the number of notes in one training samples. $^{\dagger}$ Number of features used in the training.}
\end{table}
\subsection{Study II: Effect of Different Input Features}
In order to investigate the impact of various input features, we elected five feature combinations and executed corresponding experiments on each group. These combinations are displayed in Table~\ref{tab:featurescomb}, where \textbf{D} symbolizes the usage of the deviation feature as a replacement for the original note-wise feature. 

\begin{table}[!hbt]
    \centering
    \caption{Feature combinations for study II on \textit{ID-400}}
    \begin{tabular}{cccccc}
    \hline
    \textbf{Combinations}&\textbf{C1}&\textbf{C2}&\textbf{C3}&\textbf{C4}\cite{syed2023hipi}&\textbf{C5}\\
    \textbf{Note-wise Features}&&&&&\\
    \hline
    Pitch&Yes&No&No&No&Yes\\
    Velocity&Yes&Yes&\textbf{D}&\textbf{D}&Yes\&\textbf{D}\\
    Onset time&Yes&Yes&\textbf{D}&No&Yes\&\textbf{D}\\
    Offset time&Yes&Yes&\textbf{D}&No&Yes\&\textbf{D}\\
    Duration&Yes&Yes&\textbf{D}&\textbf{D}&Yes\&\textbf{D}\\
    Inter-onset Interval (IOI)&Yes&Yes&\textbf{D}&\textbf{D}&Yes\&\textbf{D}\\
    Offset Time Duration (OTD)&Yes&Yes&\textbf{D}&No&Yes\&\textbf{D}\\
    \hline
    \end{tabular}
    \label{tab:featurescomb}
\end{table}

\textbf{C1} embodies 7 original note-wise features; \textbf{C2} omits the singular frequency-based feature, pitch, from \textbf{C1}; \textbf{C3} comprises only deviation features; \textbf{C4} replicates the same combination used in the study \cite{syed2023hipi}; while \textbf{C5} incorporates all available features. Experiments were conducted on the \textit{ID-400} dataset utilizing music segments of 1000 notes. The mean accuracy from five iterations along with the standard deviation for each feature combination experiment is detailed in Table~\ref{tab:result2}.

\begin{table}[!hbt]
    \centering
    \caption{Experiment results of Study II on \textit{ID-400} with 1000-note sequences}
    \begin{tabular}{cccc}    
    \hline
    \textbf{Combination}&\textbf{\# of Features}&\textbf{Acc. (Std.)}&\textbf{F1 (Std.)}\\
    \hline
    \textbf{C1}&7&0.729 (0.035)&0.710 (0.041)\\
    \textbf{C2}&6&0.727 (0.027)&0.724 (0.022)\\
    \textbf{C3}&6&0.711 (0.010)&0.706 (0.011)\\
    \textbf{C4}\cite{syed2023hipi}&3&0.720 (0.008)&0.719 (0.007)\\
    \textbf{C5}&13&\textbf{0.787 (0.015)}&\textbf{0.781 (0.017)}\\
    \hline
    \end{tabular}
    \label{tab:result2}
\end{table}

The results highlight negligible differences when employing either note-wise features or deviation features in isolation for training the model. Incorporation of all the features collectively yields the optimal performance, suggesting that the model is more adept at identifying a performer's style when given the full set of related features. The comparison between \textbf{C1} and \textbf{C2} groups suggests that the frequency-based feature does not make a significant contribution to the identification process. Concurrently, the outcomes provide further evidence that the combination of velocity, duration, and IOI deviations proves to be a more reliable choice when solely utilizing deviation features, as discussed in \cite{syed2023hipi}.
\subsection{Study III: Comparison between \textit{ID-400} with \textit{ID-1000}}
\begin{figure}[!hbt]
  \centering
\includegraphics[width=.9\linewidth]{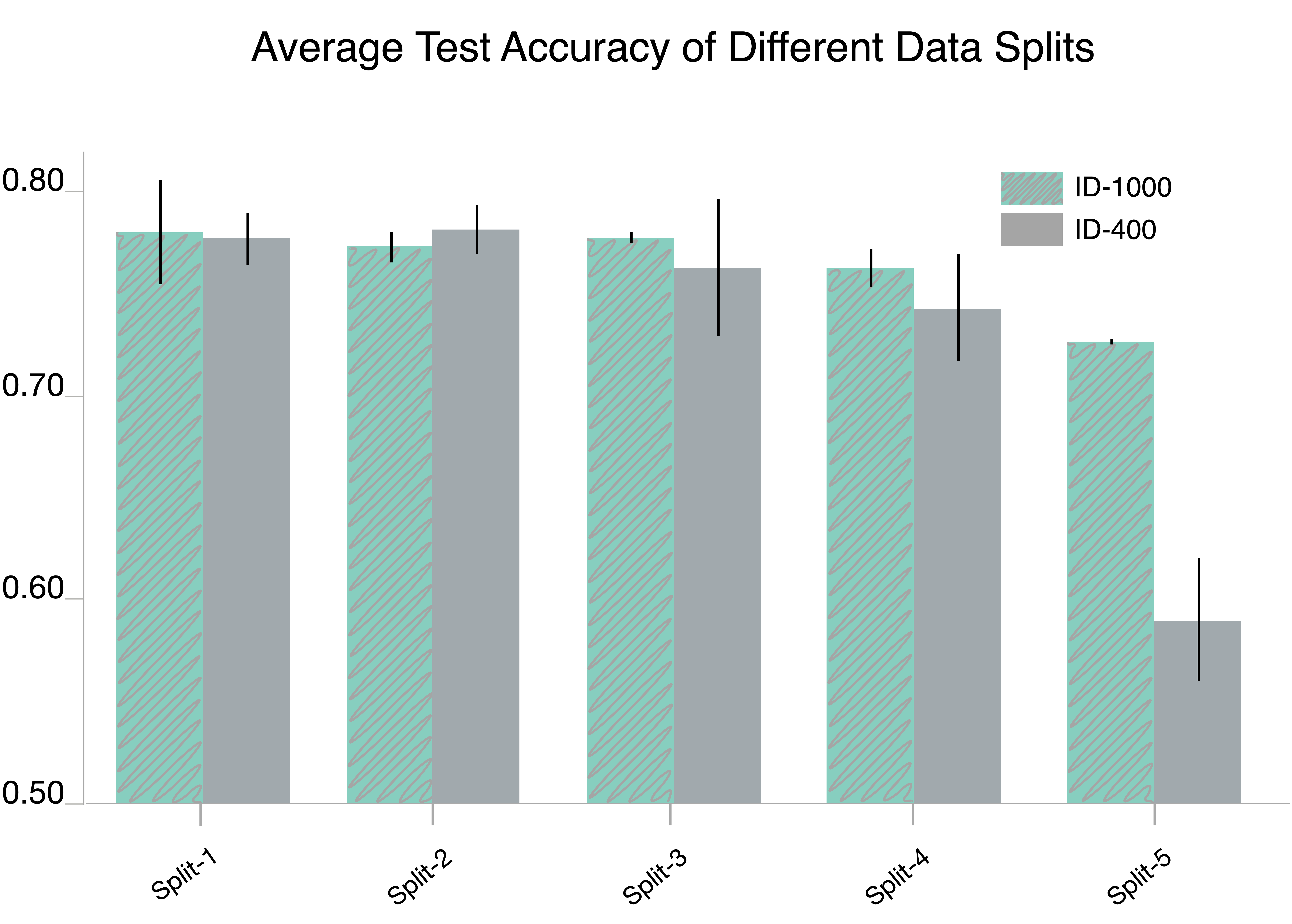}
  \caption{Comparison of the test accuracy and the standard deviation with different data splits on \textit{ID-400} and \textit{ID-1000} using segments of 1000 notes} 
   \label{fig:results}
\end{figure}

Despite the implementation of a carefully designed data splitting algorithm, the relatively small size of the subset \textit{ID-400} impedes the creation of training, testing, and validation sets that maintain similar data distributions. Employing the same algorithm, we generated five varied data splits for both the \textit{ID-400} and \textit{ID-1000} datasets, each of which underwent model testing. The Table \ref{tab:result3} presents the average test accuracy across all data splits, alongside the highest accuracy achieved by the best models on both datasets. Experiments were conducted using sequences of 1000 notes and 13 features.
\begin{table}[!hbt]
    \centering
    \caption{Comparison of the model performances on ID-400 and ID-1000 for Study III}
    \begin{tabular}{cccc}    
    \hline
    \textbf{Datasets}&\textbf{Type}&\textbf{Acc. (Std.)}&\textbf{F1-score (Std.)}\\
    \hline
    \textit{ID-400}&Best&0.799 &\textbf{0.796}\\
    \textit{ID-400}&Average&0.731 (0.077)&0.721 (0.076)\\
    \textit{ID-1000}&Best&\textbf{0.805}&0.795\\
    \textit{ID-1000}&Average&\textbf{0.766 (0.024)}&\textbf{0.754 (0.027)}\\
    \hline
    \end{tabular}
    \label{tab:result3}
\end{table}
The outcomes, as outlined in the Table \ref{tab:result3} and Fig.~\ref{fig:results}, reveal that training with \textit{ID-400} results in a model highly sensitive to alterations in data splits. In contrast, training with the more expansive \textit{ID-1000} dataset mitigates this issue, thereby enhancing the model's robustness for accurately identifying the six pianists.  

\section{Conclusion}\label{conclution}
We presented our investigation of the application of convolutional neural networks to the pianist identification task. Our proposed convolutional neural network model shows promising results in identifying virtuoso pianists. Three studies were conducted, analysing the effects of varying input sequence lengths, the utilization of diverse expressive features, and the impacts of different datasets on the model's performance. Our findings suggest that our model performs best when handling complete musical performances rather than fragments, outperforming the state-of-the-art with 85.3\% accuracy when integrating a larger set of features into the training phase. Our model uses less computational resource, leading to significant time savings during the training process compared with the state-of-art. In addition, training on our proposed larger \textit{ID-1000} dataset resulted in a model less sensitive to alterations in data splits, thereby improving the robustness in identifying the six pianists. 

Our model serves as an exemplar for embedded systems that aspire to decode and respond to nuanced musical cues. Just as voice-operated devices discern users' vocal nuances, our proposed model distinguishes pianists based on their expressive nuances. There are numerous further applications of the technology in the IoS and IoMusT contxts, including the population of music related ontologies~\cite{6517233,CERIANI2023100787,Turchet2021smo,Turchet2020iomusto} with performer identity or style related information. 

Future work could extend these findings, utilising the proposed model to develop identifiers for more pianists. Such extensions will offer a more comprehensive understanding of pianist-specific performance characteristics, and enrich the applications of the current system. It would also be beneficial to evaluate the model's generalization abilities on unseen compositions. 
\printbibliography
\end{document}